\documentclass[twocolumn,superscriptaddress, prb]{revtex4-1}
\usepackage{amsfonts}
\usepackage{amssymb}
\usepackage{amsmath}
\usepackage{amsthm}
\usepackage{dsfont}
\usepackage{mathtools}
\usepackage{stackrel}
\usepackage{color}

\usepackage[pdftex]{graphicx,hyperref}

\usepackage{hyperref}

\newcommand{\beqarr}{\begin{eqnarray}}
\newcommand{\eeqarr}{\end{eqnarray}}

\newcommand{\beq}{\begin{equation}}
\newcommand{\eeq}{\end{equation}}

\newcommand{\nogr}[1]{}

\begin{document}

\title{On the phase structure of driven quantum systems}

\author{Vedika Khemani}
\affiliation{Department of Physics, Princeton University, Princeton, NJ 08544}
\affiliation{Max-Planck-Institut f\"{u}r Physik komplexer Systeme, 01187 Dresden, Germany}

\author{Achilleas Lazarides}
\affiliation{Max-Planck-Institut f\"{u}r Physik komplexer Systeme, 01187 Dresden, Germany}

\author{Roderich Moessner}
\affiliation{Max-Planck-Institut f\"{u}r Physik komplexer Systeme, 01187 Dresden, Germany}

\author{S. L. Sondhi}
\affiliation{Department of Physics, Princeton University, Princeton, NJ 08544}
\affiliation{Max-Planck-Institut f\"{u}r Physik komplexer Systeme, 01187 Dresden, Germany}

\date{\today}

\begin{abstract}
Clean and interacting periodically-driven systems are believed to exhibit a single, trivial ``infinite-temperature" 
Floquet-ergodic phase. In contrast, here we show that their disordered Floquet many-body localized counterparts
can exhibit distinct ordered phases delineated by sharp transitions. Some of these are analogs of equilibrium states with broken symmetries and topological order, while others ---  genuinely new to the Floquet problem --- are characterized by order {\it and} non-trivial periodic dynamics. 
 We illustrate  these ideas in  
driven spin chains with Ising symmetry. 
\end{abstract}

\pacs{}
\maketitle


\noindent
{\bf Introduction:}
Extending ideas from equilibrium statistical mechanics  to the non-equilibrium setting is a topic of 
perennial interest. We consider a question in this vein: Is there a sharp 
notion of a phase in driven, interacting quantum systems? 
We find an affirmative answer for Floquet systems\cite{Shirley:1965cy,ZelDovich:1967va,Sambe:1973hi}
whose Hamiltonians  depend on time $t$  periodically, $H(t+T) = H(t)$.
Unlike in  equilibrium statistical mechanics, disorder turns out to be an essential ingredient
for stabilizing different phases; moreover, the periodic time evolution
allows for the existence (and diagnosis)
of phases without any counterparts in  equilibrium statistical mechanics. 

Naively, Floquet systems hold little promise of a complex phase structure. In systems with periodic
Hamiltonians, not even the basic concept of energy survives, being replaced instead with a 
quasi-energy defined up to arbitrary shifts of $2\pi/T$. Indeed, interacting Floquet 
systems should absorb energy indefinitely from the driving field, as suggested by  standard linear 
response reasoning wherein any nonzero frequency exhibits dissipation. 
This results in the system heating up to ``infinite temperature'', at which 
point all static and dynamic correlations become trivial and independent of starting state ---  
thus exhibiting a maximally trivial form of ergodicity\cite{D_Alessio_2014,Lazarides_2014,PonteChandran_2015}.

To get anything else requires a mechanism for energy localization 
wherein the absorption from the driving field saturates, and the long-time state of the system is sensitive to initial conditions. The current dominant belief is that translationally 
invariant interacting systems cannot generically exhibit such energy 
localization \cite{Lazarides_2014,D_Alessio_2014,PonteChandran_2015},  although 
 there are computations that suggest otherwise \cite{Prosen:1998kn,Citro_2015,chandran2015interaction}. The basic intuition is that spatially extended modes in translationally invariant systems interact with and transfer energy 
between each other.

This can be different when disorder spatially localizes the modes, with individual
modes exhibiting something like Rabi oscillations while interacting only weakly with distant modes.
While the actual situation is somewhat more involved, 
several pieces of work\cite{Lazarides_2015,PonteHuveneers_2015,abanin2014theory} 
have made a convincing case for the existence of Floquet energy 
localization
exhibiting a set of properties closely related
to those exhibited by time-independent many-body localized\cite{Basko_2006}  (MBL) systems \footnote{For a recent review see \cite{Nandkishore_2015} and references therein}.

In the following we show that such Floquet-MBL systems can
exhibit multiple phases. Some of these are driven cousins of MBL phases characterized by broken symmetries and 
topological order. Remarkably, others are genuinely new to the Floquet setting,
characterized by order {\it and} non-trivial periodic dynamics. Our analysis identifies a key feature of the Floquet problem, the existence
of Floquet eigenstates, which permits us to extend the notion of eigenstate order\cite{huse2013localization,pekker2014hilbert,vosk2014dynamical,chandran2014many,bahri2015localization} to time dependent Hamiltonians. 
Our work also builds on the discovery of topologically non-trivial Floquet single 
particle systems and recent advances in their classification \cite{Thouless:1983hb,Oka:2009kc,Jiang:2011cw,Kitagawa:2011fj,Lindner:2011ip,kitagawa2010topological,rudner2013anomalous,Nathan:2015tk,rroy2015ktheory}. As we will explain, non-trivial single particle drives 
can  yet lead to trivial many-body (MB) periodic dynamics even without interactions. Thus, the full framework of 
disorder and interactions is required for the MB problem.

In the following sections, we briefly review the Floquet formalism and describe the notion of eigenstate order before generalizing it to the Floquet setting. We then illustrate our ideas for driven Ising spin chains.
We first show that there are two Floquet phases, paramagnet (PM) and spin glass (SG), that connect smoothly to phases in the undriven systems. We then identify two {\it new} phases that do {\it not}, which we term the Floquet $0 \pi$-PM and the Floquet $\pi$-SG. Along the way we note that, reformulated as fermion problems, these yield instances of Floquet topological order. 

\noindent
{\bf Floquet Formalism:} We consider periodic, interacting Hamiltonians which are local in space.
The time dependent Schr\"odinger equation 
\beq
i {d |\psi(t)\rangle \over dt} = H(t)  |\psi(t)\rangle 
\eeq
has special solutions~\cite{Shirley:1965cy,Sambe:1973hi,ZelDovich:1967va}:
\beq
|\psi_\alpha (t)\rangle = e^{- i \epsilon_\alpha t}  |\phi_\alpha (t)\rangle  \ 
\eeq
defined by periodic states $ |\phi_\alpha (t)\rangle =  |\phi_\alpha (t+T)\rangle$ and quasi-energies $\epsilon_\alpha$ defined modulo $2\pi/T$. 

These  replace the eigenstates of the time independent problem; in them observables have periodic expectation values,
and they form a complete basis. The ``Floquet
Hamiltonian'' $H_F$ is defined via the time evolution operator over a full period, $U(T) = e^{-i H_F T}$. The  $ |\phi_\alpha (0) \rangle$ are eigenstates of both $U(T)$ and $H_F$, with eigenvalues $e^{- i \epsilon_\alpha T}$ and $\epsilon_\alpha$ respectively.

The question of energy localization relates to the action of $U(nT)=(U(T))^n$ as $n \rightarrow \infty$. An equivalent formulation of Floquet-MBL is that there exists an $H_F$ which is local and exhibits the generic properties of any fully MBL Hamiltonian, such as a full set of conserved local operators\cite{Serbyn:2013ug,Imbrie:2014vo,Huse:2014co,Ros:2015ib,Chandran:2015cw} ($l$-bits), area law eigenstates and failure of the eigenstate thermalization hypothesis (ETH)\cite{Pal:2010gr}.

\noindent
{\bf Eigenstate Order:} Traditionally, phase transitions at non-zero energy densities are considered in the framework of quantum statistical mechanics, signaled by singularities in thermodynamic functions or observables computed in the $T > 0$ Gibbs state. Work on MBL has led to the realization that this viewpoint is too restrictive\cite{huse2013localization,pekker2014hilbert,vosk2014dynamical,chandran2014many,bahri2015localization}---instead the MB eigenstates and eigenvalues can directly exhibit singular changes  as a parameter is varied. Such transitions have been termed eigenstate phase transitions. This distinction is irrelevant for Hamiltonians obeying ETH, but when ETH fails it becomes important. The passage from ergodicity to localization is an example of an eigenstate transition undetectable by standard ensembles\cite{Pal:2010gr,OganesyanHuse,vznidarivc2008many}---it can take place at $T=\infty$ and yet the individual MB eigenstates are sharply different in their entanglement properties, with the eigenvalue distributions exhibiting different statistics. Moreover, eigenstate phase transitions can take place between two ETH violating phases\cite{huse2013localization} and they may even involve a singular rearrangement of the eigenvalues alone. 

We now generalize this to the Floquet-MBL regime, leading us to find multiple ordered phases whose existence can be detected in the Floquet MB eigensystem. In general, this will require examination of the full periodic solutions $ |\psi_\alpha (t)\rangle$ for sharply different characteristics. We will demonstrate our results in the simple interacting setting of a one dimensional disordered spin chain with Ising symmetry,
\begin{align}
H = \sum_i J_i \sigma_i^x \sigma_{i+1}^x  + \sum_i h_i \sigma_i^z + J_z \sum_i \sigma_i^z\sigma_{i+1}^z. 
 \end{align}
Carrying out a Jordan-Wigner transformation on only the first two terms gives a p-wave superconducting free-fermion model, whereas the final term is a density-density interaction in the fermion language. The paramagnetic and symmetry-broken ferromagnetic phases of the Ising model are related by a well-known duality. 

\begin{figure}[htbp]
\includegraphics[width = \columnwidth]{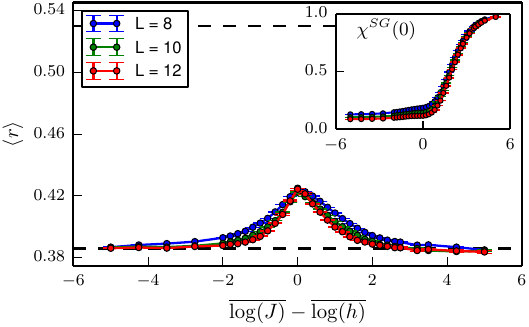}
\caption{Disorder averaged level statistics $\langle r \rangle$ of $H_F$ for the driven, disordered Ising model (\ref{eq:SGDrive}). $\langle r \rangle$ approaches the Poisson limit of .386 with increasing $L$ deep in the PM and SG phases, showing that these remain well localized. There is a peak in $\langle r \rangle$ near the non-interacting critical point at $\overline{\log J } = \overline{\log h}$ indicating partial delocalization, although the value still remains well below the COE value of .527. (inset): The SG diagnostic $\chi^{SG}$ defined in (\ref{eq:SG}) goes to 0 in the PM and approaches a non-zero value in the SG phase. All data is averaged over $2000-10^5$ samples depending on $L$. }
\label{Fig:DrivenIsing}
\end{figure}

\noindent
{\bf Floquet Paramagnet and Spin Glass:} We begin with the two phases that {\it do} exist in undriven systems and demonstrate
 the stability of these to being (not too strongly) driven. 
Starting with the non-interacting limit, $J_z=0$, we choose the $J_i$ and the $h_i$ to be log-normally distributed with a tunable mean $\overline{\log(J_i)}\equiv
\overline{\log J }$, fixed $\overline{\log(h_i)} \equiv \overline{\log h} =0$ and two fixed and equal standard deviations $\delta \log(h_i) = 
\delta{\log(J_i)} = 1$. Work on random, non-interacting Ising models culminating in Ref.~\onlinecite{fisher1995critical}
finds a ground state phase diagram which is a paramagnet for 
$\overline{\log J} < \overline{\log h}$ and a $Z_2$ breaking ferromagnet for $\overline{\log J} >  \overline{\log h}$, separated 
by an infinite disorder fixed point at $\overline{\log J } = \overline{\log h}$. The work on eigenstate order has shown 
that, with disorder and localization, both phases exist at {\it all} energies with the symmetry-breaking phase exhibiting 
SG order in individual eigenstates instead of ferromagnetism. The eigenstates are also eigenstates of parity  $P = \prod \sigma_i^z$, and deep in the PM phase, they (roughly) look like frozen spins along the $z$ direction $|\uparrow\downarrow\downarrow\cdots \uparrow\rangle$ while deep in the SG phase they look like global superposition/cat states with spins in the $x$ direction with frozen domain walls $|\pm \rangle = \frac{1}{\sqrt{2}}(|\rightarrow \leftarrow \rightarrow \cdots \rightarrow \rangle \pm |\leftarrow \rightarrow\leftarrow \cdots \leftarrow \rangle)$. 

With weak interactions, $0 < J_z \ll 1$, the strongly localized PM and SG phases remain MB localized\cite{Pekker:2014bj,Kjall:2014bd}. 
The fate of the SG-PM transition is more sensitive to the inclusion of interactions. It was suggested that it would remain localized\cite{huse2013localization}  and exhibit the same scaling as the non-interacting fixed 
point\cite{Pekker:2014bj}; we comment on the analogous question in the Floquet setting below.

We consider a periodic binary drive---computationally much simpler than a monochromatic modulation---switching between 
two static $H$s with $\overline {\log J }$ differing by 1: 
\begin{align}
H(t) = \sum_i f_{\rm s} (t) \;J_i \sigma_i^x \sigma_{i+1}^x  + \sum_i h_i \sigma_i^z + J_z \sum_i \sigma_i^z\sigma_{i+1}^z ,
\nonumber\\
f_{\rm s} (t) = \begin{cases} 1 & \quad \text{if } 0 \le t < \frac{T}{4}  \, \text{   or   } \, \frac{3T}{4}< t \leq T \\ e & \quad \text{if } \frac{T}{4} \leq t \leq\frac {3T}{4} \\ \end{cases}   .
\label{eq:SGDrive}
\end{align}
We set $J_z=0.1$ in the following. 
For $-1 \le \overline{ \log J} \le 0$ the drive straddles the undriven phase transition, up to  small corrections to its location due to the interaction.

Drives consistent with  Floquet localization require  both small interactions and not too small frequencies. 
We arrange the latter by defining, for each set of $(\overline{\log J}$, $\overline{\log h})$ parameters, an 
effective ``single-particle bandwidth'',  $W = \max (\sigma_{J}, \sigma_{h})$, where $\sigma_h$ and $\sigma_J$ are the
 standard deviations of $h_i$ and $J_i$ determined from the underlying log-normal distributions. 
 The period is then defined by $\omega = 2\pi/T = 2W$. This choice ensures a roughly constant ratio 
 of $\omega/W$ for different $\overline{\log J} - \overline{\log h}$ values and thus isolates the effect of tuning the means through the phase diagram. 
 
 The lowest frequency in our drives is bigger than the estimated single particle
 bandwidth but much smaller than the MB bandwidth, so that localization is not a foregone conclusion. 
 In Fig.~\ref{Fig:DrivenIsing} we characterize the quasi-energy spectrum $\epsilon_{n} \in [0, 2\pi)$ using the level statistics of 
$H_F$. We define quasi-energy gaps by $\delta_n = \epsilon_{n+1} - \epsilon_{n}$ 
and the level-statistics ratio $r = \min(\delta_n, \delta_{n+1})/\max(\delta_n, \delta_{n+1})$. 
Away from the critical region which---given the weak interactions and large frequency of the drive---is close to the undriven, non-interacting transition point $\overline{\log J} = \overline{\log h}$, the disorder averaged $\langle r \rangle$ approaches the Poisson limit of $.386$ with increasing system size $L$, signaling a lack of level repulsion and hence MBL. In the interacting critical region we find a peak in $\langle r \rangle $ which does 
not grow with system size and is much less than the delocalized COE value of $.527$; we return to this below.

With localization established, we turn to distinguishing the phases. Consider a pair of $Z_2$ invariant correlators
(with $A=x$ or $y$)
\begin{align} \label{eq:corr}
C_{AA}^{\alpha} (ij;t) &= \langle \phi_\alpha(t)| \sigma_i^A \sigma_j^A | \phi_\alpha(t)  \rangle 
\end{align}
for $i-j \gg 1$ in any given Floquet eigenstate. We find that for $\overline{\log J} < \overline{\log h}$ both correlators vanish
with increasing system size $L$ at all $t$, signaling a PM. For $\overline{\log J} > \overline{\log h}$, 
both are generically non-zero, though of random sign varying with eigenstate and location, signaling SG order.  
For the SG, our parameters give $|C_{xx}^{\alpha}(ij;t)| \gg |C_{yy}^{\alpha}(ij;t)|$ at all $t$ although the more general 
signature is that $|C_{xx}^{\alpha}(ij;t)|$ and $|C_{yy}^{\alpha}(ij;t)|$ do not cross 
for $0<t<T$. For our parameters, it suffices to compute
\beq
\chi_\alpha^{SG}(t)  = \frac{1}{L^2}\sum_{i,j = 1}^L ( \langle \phi_\alpha(t)| \sigma_i^x \sigma_j^x | \phi_\alpha(t)  \rangle )^2
\label{eq:SG}
\eeq
for $t=0$. In Fig. \ref{Fig:DrivenIsing} (inset) we plot the disorder averaged $\chi_\alpha^{SG}(0)$; the trend with system-size indicates that $\chi^{SG}(t) > 0$ in the spin-glass and $\chi^{SG}(t) \rightarrow 0$ 
in the paramagnet. 

Three comments are in order. First, recall that it would be sufficient to establish 
the existence of the Floquet PM---the SG can be obtained by duality\footnote{In the tails of the SG phase, $\overline{\log(J)} - \overline{\log(h)} \gtrsim 2$, the values of the couplings $J$ are much bigger than the fixed interaction strength $J_z = 0.1$ so the problem effectively looks non-interacting. The PM does not suffer from this problem, and the strongly localized, interacting SG can be simply obtained by dualizing the PM.}. 
Second, in chains with uniform couplings, both spin and dual spin order vanish in all but one of the Floquet eigenstates (the notion of the ``ground-state'' is not well defined in a Floquet system) even without interactions---this is the Landau-Peierls 
prohibition against discrete symmetry breaking in disguise. 
Localization is essential to avoid this. Third, the $|\pm\rangle$ MB Floquet eigenstates in the localized SG phase come in conjugate, almost degenerate pairs with different parity but with similar domain wall configurations. In the fermionic formulation of the problem, the PM is topologically trivial while the SG is non-trivial. The non-interacting SG phase has zero energy edge Majorana modes in open chains, and the two-fold degeneracy of the many-body SG spectrum (in this language) stems from the occupation/unoccupation of the bilocal Dirac mode formed from the edge Majoranas. With interactions, the edge mode remains coherent only in the MBL setting\cite{chandran2014many,bahri2015localization}. Thus, the degenerate Floquet eigenstates can be connected by either (i) spectrum-generating operators localized near the edges which toggle the state of the coherent edge mode (fermionic language) or (ii)  any spin operator that flips the parity of the eigenstates. Concretely, the spectral function of $\sigma_i^+$, the spin raising operator on any site $i$, in the Floquet eigenbasis 
\begin{equation}
\mathcal{A}(\omega) = \frac{1}{2^L}\sum_{\alpha \beta} \langle \phi_\alpha(0) | \sigma_i^+ | \phi_\beta(0) \rangle \delta (\omega-(\epsilon_\alpha -\epsilon_\beta)) 
\label{eq:spec}
\end{equation} 
is a delta function peaked at $\omega = 0$ (this phase will hence also be labeled the `$0$' phase below). 
Finally, we note that the SG displays long-range string order in all eigenstates regardless of boundary conditions. 
Without disorder, the string order vanishes even in the many body eigenstates of free fermion chains---despite the non-trivial momentum space topology present in their Hamiltonians.

\noindent
{\bf Paramagnet-Spin Glass Phase transition:}  In the non-interacting problem we have strong evidence that the 
infinite disorder fixed point continues to control the physics. 
We have examined $H_F$ and we find that all its eigenstates are localized even at the transition, and its structure differs from 
the canonical strong-disorder renormalization group form\cite{fisher1995critical} by short ranged, irrelevant, terms. 
The ultimate fate of the critical region in the interacting driven problem is an interesting open question\cite{DrivenCP}, 
but we note for now that our data on $\langle r \rangle$ suggests a partially delocalized interacting critical point.

\begin{figure*}[htbp]
\includegraphics[width = \textwidth]{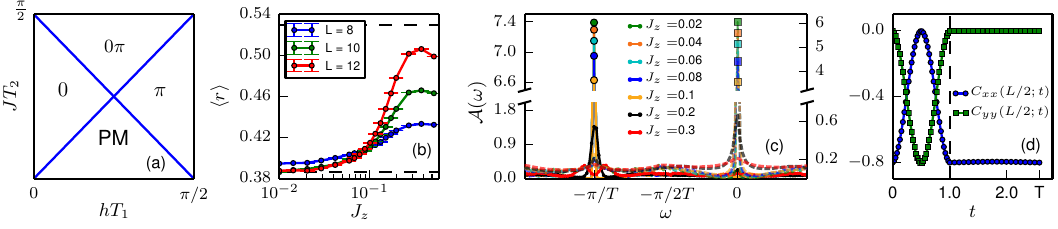}
\caption{ (a) Phase diagram for the binary Ising drive, Eq.~(\ref{eq:piDrive}) without interactions ($J_z = 0$) and disorder. (b) Level Statistics $\langle r \rangle$ of $H_F$  in the $\pi$ phase with parameters defined in the text and disorder averaged over 2000-100,000 realizations for different $L$s. $H_F$ is localized for interaction strengths $J_z \lesssim 0.1$. 
(c) Disorder averaged spectral function $\mathcal{A}(\omega)$ defined in Eq.~(\ref{eq:spec}). Solid (dashed) lines are for the $\pi$-SG  ($0$-SG)  phase showing a delta function peak at $\omega = \pi/T\; (0)$ for small interaction strengths which disappears as the interaction is increased. (d) Time dependence of the $C_{xx}$ and $C_{yy}$ correlators defined in Eq.~(\ref{eq:corr}) over one period for an eigenstate in the $\pi$ phase; $L = 10$, $T_1=1$ and $J_z = 0.04$. The crossings are robust in the $\pi$ phase. }
\label{Fig:DrivenPi}
\end{figure*}

\noindent
{\bf $\pi$ Spin Glass and $0\pi$ Paramagnet:} We now present two new Ising phases which exist 
only in the driven system---the $\pi$-SG and the $0\pi$-PM. Existing work on the band topology of translationally 
invariant $Z_2$ symmetric free-fermion chains\cite{Jiang:2011cw, Bastidas2012, Thakurathi:2013dt} has shown that the Floquet eigenmodes for such chains with {\it open} boundary conditions can exhibit edge Majorana
modes with $\epsilon_\alpha = \pi/T$ in addition to the better known edge modes with  $\epsilon_\alpha = 0$.  In the MBL setting in the `$\pi$' phase, the MB Floquet eigenstates are long-range ordered and come in $|\pm \rangle$ cat pairs separated by quasienergy $\pi/T$.  These can again be connected by either spectrum generating 
operators localized near the two edges (fermion language) or by local parity odd operators (spin language). Thus, the spectral function $\mathcal{A}(\omega)$ \eqref{eq:spec} now shows a delta function peak at $\omega = \pi/T$. 

We now establish these phases for the binary periodic drive
\beq
H(t) = \begin{cases} H_z & \quad \text{if } 0 \le t < T_1 \\ H_x & \quad \text{if } T_1 \le t < T=T_1+T_2 \\ \end{cases} 
\label{eq:piDrive}
\eeq

\begin{align*}
  H_z &= \sum_{i=1}^{L} h_i \sigma_i^z + \sum_{i=1}^{L-1} J_z \sigma_i^z \sigma_{i+1}^z, \\
  H_x &= \sum_{i=1}^{L-1} J_i \sigma_i^x \sigma_{i+1}^x +  J_z \sigma_i^z \sigma_{i+1}^z  \ .
\end{align*}
Fig.~\ref{Fig:DrivenPi}(a) shows the uniform, non-interacting phase diagram with the four possible driven Ising phases. The phases labeled `$0$' and ` $\pi$'  have edge Majorana modes at quasienergies $0$ and $\pi/T$ respectively.  With disorder and localization, these phases display long-range SG eigenstate order in the correlators (\ref{eq:corr}) for both $A = x, y$. Moreover, in the $\pi$-SG phase, the time dependence of the $C_{xx}$ and $C_{yy}$ correlators over the period is non-trivially correlated: their magnitudes must cross twice during a period. Thus, in this phase, the axis of SG order rotates by an angle $\pi$ about the $z$-axis during the period which can be intuitively understood by thinking semi-classically about the drive \eqref{eq:piDrive} at the extremal boundaries of the phase diagram shown in Fig.~\ref{Fig:DrivenPi}(a). 
A referee has noted that this sign reversal of the order parameter and thus doubling of the period (also found previously in \cite{AC_TC}), provides a potential Floquet 
realization of a time crystal \cite{WilczekTC, OshikawaTC}. As before, without localization, only one of the Floquet eigenstates(analogous to the ground state) will display long-range order in the $0, \pi$ phases. The other two phases, labeled PM and $0\pi$, have no long-range order and are respectively dual to the $0$ and $\pi$ phases. 

We now turn to numerically identifying the localized $\pi$ phase with disorder and interactions. We pick $T=1, T_2 = \pi/2$ and $h_i T_1$ uniformly from the interval $(1.512, 1.551)$ and $J_i T_2$ from  $(0.393,  1.492)$, so that all pairs of values $(h_i T_1, J_j T_2)$ lie in the $\pi/T$ Majorana region of the free 
uniform chains.  We have confirmed that the free fermion disordered drive  exhibits $\pi/T$ Majoranas for open chains while all other modes are localized in the bulk. In Fig 2b we examine stability to interactions via $\langle r \rangle$ and clearly observe a transition around $J_z \approx 0.1$, with the small $J_z$ regime being the MBL phase we seek. 

Fig. 2c shows the appearance  of the $\pi/T$  peak in the disorder averaged spectral function \eqref{eq:spec} for system size $L = 10$ as the delocalization transition is crossed by decreasing the interaction strength.  In contrast, the dashed lines show the spectral function for a similar drive in the $0$-SG phase, clearly showing a peak at $\omega = 0$. 
Finally,  Fig 2d displays the anticipated time dependence of the SG order in the $C_{xx}$ and $C_{yy}$ correlators in a single eigenstate of the interacting system with $J_z = 0.04$ and $L = 10$. The crossings in the correlator within the period are robust in the $\pi$-SG phase and topologically distinct from the correlators in the $0$-SG phase where there are no crossings. We emphasize that this is a true bulk diagnostic of the phase which, unlike the presence of an edge mode, is insensitive to boundary conditions. We also note that the non-trivial spin dynamics captured by it {\it cannot} be obtained without localization. 
 
Finally we turn to the $0 \pi$-PM which is dual to the $\pi$-SG.  This is an SPT phase with no bulk long-range order, but with coherent edge states. In the fermionic language, there are now two Majorana modes at each edge, one at quasienergy $0$ and the other at $\pi/T$ and thus MB spectrum is paired into conjugate sets of four MB states---two degenerate pairs of states separated by quasienergy $\pi/T$. The eigenstates in this phase do \emph{not} look like global superposition states and the spectral function of bulk spin operators shows no structure. On the other hand, spectral functions of edge operators which toggle the state of the edge modes show a peak at both $0$ and $\pi/T$. 

\noindent
{\bf Summary and open questions:} 
We have shown that MBL Floquet systems exhibit sharply defined phases 
bounded by parameter surfaces across which  properties of their Floquet eigensystems change in a singular fashion. 
These phases include the trivial Floquet-ergodic phase and multiple non-trivial non-ergodic phases 
exhibiting various forms of ordering and dynamics, some of which are entirely new to Floquet systems. 
The net result is something quite striking given the contentious history of non-equilibrium statistical mechanics.
Indeed, it is quite likely that Floquet systems constitute the maximal class for which such a definition of phase structure
is possible; with generic time dependences it would not be surprising if heating to infinite temperatures is the inevitable
result. Going forward we anticipate a more systematic search for Floquet phases and a better 
understanding of their phase diagrams. In this context we note two studies that include disorder, but not interactions, in Floquet systems \cite{titum2015disorder,titum2015anomalous}. It should also be possible to observe the new localized phases by the methods of Ref. \onlinecite{schreiber2015observation}.

\noindent
{\bf Acknowledgements:} We are especially grateful to Curt von Keyserlingk and Rahul Roy for many helpful discussions on topology in Floquet systems and to Anushya Chandran, Arnab Das and David Huse for discussions on Floquet and MBL matters.  This work was supported by NSF Grant No. 1311781 and the John Templeton Foundation (VK and SLS) and the Alexander von Humboldt Foundation and the German Science Foundation (DFG) via the Gottfried Wilhelm Leibniz Prize Programme at MPI-PKS (SLS) and the SFB 1143.

\bibliography{MBLFloquetSymmetry}

\end{document}